# A Stochastic Model for Quantitative Security

# Analyses of Networked Systems

Xiaohu Li, Paul Parker, and Shouhuai Xu

Department of Computer Science, University of Texas at San Antonio

{xli,pparker,shxu}@cs.utsa.edu

Corresponding author: Shouhuai Xu (shxu@cs.utsa.edu)

**Abstract**

Traditional security analyses are often geared towards cryptographic primitives or protocols. Although such analyses are necessary, they cannot address a defender's need for insight into *which aspects of a networked system having a significant impact on its security, and how to tune its configurations or parameters so as to improve security*. This question is known to be notoriously difficult to answer, and the state-of-the-art is that we know little about it. Towards ultimately addressing this question, this paper presents a stochastic model for quantifying security of networked systems. The resulting model captures two aspects of a networked system: (1) the strength of deployed security mechanisms such as intrusion detection systems, and (2) the underlying *vulnerability graph*, which reflects how attacks may proceed. The resulting model brings the following insights: (1) How should a defender "tune" system configurations (e.g., network topology) so as to improve security? (2) How should a defender "tune" system parameters (e.g., by upgrading which security mechanisms) so as to improve security? (3) Under what conditions is the steady-state number of compromised entities of interest below a given threshold with a high probability? Simulation studies are conducted to confirm the analytic results, and to show the tightness of the bounds of certain important metric that cannot be resolved analytically.

## I. INTRODUCTION

Traditional security analyses are often geared towards cryptographic primitives (e.g., how one should pad a message before encrypting it using the RSA function) or protocols (e.g., how a password-based authentication protocol should operate so that it is immune to the off-line dictionary attack). Although such analyses are necessary, they do not provide much insight into answering an equally important, if not more important, question: *which aspects of a networked system having a significant impact on its security, and how to tune system configurations or parameters so as to improve security*. The state-of-the-art is that we know little about the answer to this question. Nevertheless, resolving this question would represent a significant step towards fulfilling full-fledged quantitative security analyses of networked systems – a notoriously difficult problem but of important practical value (cf. [5], [20]).

## A. Our Contributions

Towards ultimately addressing the above question, this paper presents a stochastic model for analyzing some aspects of networked system security. Our modeling approach can be characterized as follows.

- First, a *vulnerability graph* is used to abstract a networked system, where a vertex my represent a vulnerability or a system with possibly multiple vulnerabilities, and an arc or edge captures the relation that the exploitation of one vulnerability could lead to the exploitation of the other. Such graphs can be obtained, for example, by combining the output of vulnerability scanners and the systems configurations. (There have been some works on generating vulnerability-like graphs, such as [9], [21], [27], [12], [3], [8].)

- Then, a stochastic process (specifically, a *renewal process*) is used to describe attacks over the vulnerability graph. This allows us to capture quantitative security of a networked system via "*the probability that a* randomly picked node *is compromised when the system enters its steady state*," which facilitates analyses that lead to the following useful insights.

  * The model shows that degree distribution of the vulnerability graph has an important impact on security of a networked system. We offer a generic method to judge any two degree distributions, if comparable, from a security perspective. We also evaluate three representative types of degree distributions, namely regular graphs, random graphs, and power-law graphs. For two degree distributions of the same type, we give a method to determine which distribution will leads to a more secure system. In the more challenging case where two degree distributions are different (e.g., power-law graph vs. random graph), we manage to give a method to determine which degree distribution leads to a more secure system. These results would allow a defender to "tune" software configurations so as to improve security (e.g., by isolating some machines from others via firewalls).

  * For a given vulnerability graph, our model is able to show how a defender should "tune" system parameters (e.g., by upgrading certain security mechanisms) so as to enhance security. This is especially important when there is a constraint on the incurred financial budget (i.e., the defender needs to decides which security mechanisms should be upgraded).

      * We give a sufficient condition as well as a necessary condition, under which the number of compromised vertices are below a given threshold (e.g., one third) with a high probability. This has an important value because it offers a method to ensure that the system has no more than the threshold number of compromised vertices. This insight may be particularly relevant in distributed computing, where it is often *assumed* that there is an upper bound on the number of faulty machines [26] (otherwise, many computing tasks are not possible).

Simulation studies are conducted to confirm the analytic results, and to show the tightness of the bounds of certain important metric that cannot be resolved analytically.

*Remark 1.1:* Our model can handle arbitrarily large systems (e.g., consisting of millions of nodes), due to the abstraction of vulnerability graphs via their degree distributions.

**Organization**: The rest of the paper is organized as follows. In Section II we present our model. In Section III we explore the impact of topologies of vulnerability graphs on system security. In Section IV we explore the effect of tuning the system parameters other than topology. In Section V we explore conditions under which the steady-state number of compromised vertices is below a threshold. In Section VI we report the results of our simulation studies. We discuss related work in Section VII, and conclude the paper in Section VIII. For the purpose of better readability, we defer most of the proofs to the Appendix.

## II. MODEL

A vulnerability graph is finite graph $G = (V, E)$, where $V$ is the vertex or node set with $|V| > 0$, and $E$ is the edge set with $E \neq \emptyset$. A vertex or node $v \in V$ represents a vulnerability and an edge $(u, v) \in E$ means that the exploitation of vulnerability $u$ can lead to the exploitation of vulnerability $v$, and vice versa. (In the case of directed graphs, only one direction is allowed. Since all the results in this paper are equally applicable to both directed and undirected graphs, we assume that they are undirected.) For a randomly picked vertex $v \in V$, denote by

$$\Pr[D = i] = p_i \text{ for } i = 0, 1, 2, \ldots,$$

where $D$ is the random variable indicating the number of a randomly picked vertex's neighbors. In practice, $i = 1, 2, \ldots, |V| - 1$, meaning that the distribution is truncated; this does not impose any problem for large $|V|$.

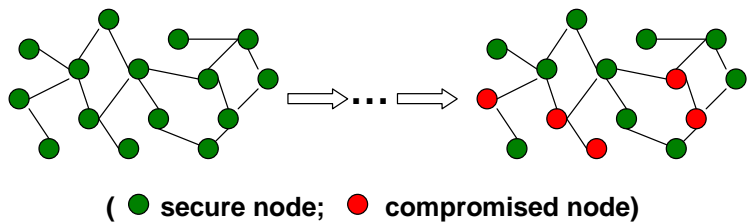

( ● secure node;   ● compromised node)

Fig. 1.  State evolution of networked system with respect to time

Figure 1 illustrates an example scenario of system state evolution. Initially, no vertex (or node) may be compromised. As time goes by, some nodes may get compromised, and the compromise may be detected and then fixed. This process may repeat for many times. An important issue here is that we need to capture the impact of the states of a node's neighbors on its own state because a node may get compromised through an attack that is launched from one or multiple neighbors.

In order to capture the strength of security mechanisms deployed by the corresponding machines, we classify them into two categories: *attack-prevention mechanisms* such as virus or firewall filters, and *attack-detection-and-recovery mechanisms* such as intrusion detection systems. An attack succeeds if it can penetrate attack-prevention mechanisms. A successful attack is typically detected after a delay in time, and then dealt with some appropriate countermeasures.

We consider a continuous time model with $t \geq 0$. At any point in time, $v \in V$ is either secure or compromised. We make the following assumptions. (1) The time, $X_1$, that a secure vertex, which has no compromised neighbors, gets compromised follows the exponential distribution with rate $\alpha$. (2) The time, $X_{2,i}$, that a secure vertex gets compromised because of its $i^{th}$ compromised neighbor, follows the exponential distribution with rate $\gamma$. (3) The time, $Y_1$, that a compromised vertex becomes secure again because the compromise is detected, follows the exponential distribution with rate $\beta$. (4) The time, $Y_2$, that a compromised vertex becomes secure again because of any other reason, follows the exponential distribution with rate $\eta$. In addition, we assume that all the random variables mentioned above, namely the $X_1, Y_1, Y_2$, and all the $X_{2,i}$, are mutually independent. This independence assumption by no means suggests that the event of one vertex getting compromised is independent of the event of another vertex getting compromised. In contrast, our model actually captures the fact that one vertex getting compromised indeed increases the likelihood of its neighbors getting compromised (i.e., the events that vertices get compromised are indeed correlated).

*Remark 2.1:* Notice that we differentiate the above (1) and (2) because some vertices may get compromised directly by the attacker, and some vertices may get compromised by some other compromised vertices (i.e., the attacker's "stepping stones"). We differentiate (3) and (4) because some compromised vertices may become secure just because they adopt a software patch, which becomes available only after the detection of some other vertices having been compromised. If the above differentiations are not important, then one can simply set, say, $\Pr[Y_2 = \infty] = 1$ (or $\eta = 0$).

**Notations**. Our model utilizes the concept of *stochastic order* [30] to compare two random variables. Specifically, for random variables $R_1$ and $R_2$, "$R_1 \preceq_{st} R_2$" means $\Pr[R_1 > x] \leq \Pr[R_2 > x]$ for any $x$. Naturally, we can say that $R_2$ is a "stochastically upper bound" of $R_1$. Below we summarize the main notations used in this paper.

| | |
|---|---|
| $\alpha$ | the rate a secure vertex with no compromised neighbors gets compromised |
| $\gamma$ | the rate a secure vertex gets compromised because of a compromised neighbor |
| $\beta$ | the rate a compromised vertex becomes secure because the attack is detected |
| $\eta$ | the rate a compromised vertex becomes secure because of any other reason |
| $D$ | a random variable indicating the number of a randomly picked vertex's neighbors with $\mu = \mathsf{E}[D]$ |
| $K$ | a random variable indicating the number of a randomly picked vertex's compromised neighbors |
| $q$ | the probability a randomly picked vertex is compromised when the system enters steady state (i.e., $q = \sum_{v \in V} q_v / |V|$, where $q_v$ is the steady-state probability that $v$ is compromised.) |
| $p$ | the probability a randomly picked vertex is secure when the system enters steady state (i.e., $p = 1 - q$) |
| $C_t$ | the total number of compromised vertices at time $t$ where $\mathsf{E}[C_{t \to \infty}] = q \cdot |V|$ |

*Remark 2.2:* In practice, the parameters may be derived from history data. It is also possible that such derived parameters may need to be adjusted by human security experts (as an analogy, we mention that even though we have very advanced medical equipments such as Computed

Tomography (CT) systems, the role of human doctors can never be overestimated). Yet another scenario is that the model can be used by security administrators to conduct "what if" analyses based on assigned parameters (e.g., the sets of parameters they care most).

## A. General Result

We model the evolution of a randomly picked vertex through a *renewal process*. Each cycle of the renewal process is composed of the time interval $X$ corresponding to the secure state, and the time interval $Y$ corresponding to the compromised state. The *security metric* we use is $q$, namely "the probability that a randomly picked vertex is compromised when the system enters its steady state." Intuitively, the smaller $q$ is, the more secure the system is. As a corollary, the expected number of compromised vertices when the system enters its steady state is $\mathsf{E}[C_{t\to\infty}] = q \cdot |V|$, where $V$ is the vertex set of the system or graph, and $C_t$ is a random variable indicating the number of compromised vertices at time $t$.

*Theorem 2.3:* Given the parameters and assumptions specified above, in the long run, the probability $q$ that a randomly picked vertex is compromised satisfies the following equation:

$$\frac{1}{q} - 1 = h(\alpha, \beta, \gamma, \eta, D; q), \tag{II.1}$$

where $h(\alpha, \beta, \gamma, \eta, D; x)$ is a function of variable $x$ such that

$$h(\alpha, \beta, \gamma, \eta, D; q) = \mathsf{E}\left[\frac{\beta + \eta}{\alpha + \gamma K}\right],$$

where given $D = d$, $K$ follows the binomial distribution with parameter $(d, q)$.

*Proof:* First, denote by $X_2 = \min\{X_{2,1}, \cdots, X_{2,K}\}$ the time for $v$ to get compromised because of its compromised neighbors. Then, the time that a vertex is secure in a cycle is

$$X = \min\{X_1, X_2\} = \min\{X_1, \min\{X_{2,1}, \cdots, X_{2,K}\}\} = \min\{X_1, X_{2,1}, \cdots, X_{2,K}\}.$$

Given $K = k$, $X_2$ follows the exponential distribution with rate $k\gamma$. Therefore, $X$ is also exponentially distributed with rate $\alpha + k\gamma$. By duplicate expectation [29], it holds that

$$
\begin{aligned}
\mathsf{E}[X] &= \sum_{d=1}^{|V|-1} \mathsf{E}[X|D=d] \cdot \Pr[D=d] \\
&= \sum_{d=1}^{|V|-1} \mathsf{E}[\min\{X_1, X_{2,1}, \cdots, X_{2,K}\}|D=d] \cdot \Pr[D=d] \\
&= \sum_{d=1}^{|V|-1} \left[ \sum_{k=0}^{d} \mathsf{E}[\min\{X_1, X_{2,1}, \cdots, X_{2,K}\}|K=k, D=d] \cdot \Pr[K=k|D=d] \right] \cdot \Pr[D=d] \\
&= \sum_{d=1}^{|V|-1} \left\{ \sum_{k=0}^{d} \left[ \frac{1}{\alpha + k\gamma} \cdot \binom{d}{k} q^k (1-q)^{d-k} \right] \cdot \Pr[D=d] \right\} \\
&= \sum_{d=1}^{|V|-1} \mathsf{E}\left[ \frac{1}{\alpha + \gamma K} \middle| D = d \right] \cdot \Pr[D=d] \\
&= \mathsf{E}\left[ \frac{1}{\alpha + \gamma K} \right].
\end{aligned}
$$

Second, the length of time that a vertex is compromised in a cycle $Y = \min\{Y_1, Y_2\}$ follows the exponential distribution with rate $\beta + \eta$, it is clear that

$$\mathsf{E}[Y] = \mathsf{E}[\min\{Y_1, Y_2\}] = \frac{1}{\beta + \eta}.$$

Finally, by Blackwell's theorem [29], $p$, the probability that a randomly picked vertex is secure when the system enters its steady state is

$$p = \frac{\mathsf{E}[\min\{X_1, X_2\}]}{\mathsf{E}[\min\{X_1, X_2\}] + \mathsf{E}[\min\{Y_1, Y_2\}]} = \frac{\mathsf{E}\left[ \dfrac{1}{\alpha + \gamma K} \right]}{\dfrac{1}{\beta + \eta} + \mathsf{E}\left[ \dfrac{1}{\alpha + \gamma K} \right]},$$

and thus

$$q = 1 - p = \frac{\dfrac{1}{\beta + \eta}}{\dfrac{1}{\beta + \eta} + \mathsf{E}\left[ \dfrac{1}{\alpha + \gamma K} \right]}. \tag{II.2}$$

Eq. (II.1) follows immediately.                                                                                    ∎

*Remark 2.4:* It can be seen from the proof of the above theorem that the derivation of Eq. (II.1) takes advantage of a structure similar to what is known as "series system reliability model" (cf., e.g., [32]). Moreover, we also used, essentially, security notions such as "mean time to secure failure" and "mean time to security repair" proposed in [15], but only as intermediate indices/variables (because our main index is $q$). We stress, however, a crucial issue here that we are able to accommodate graph structure through the degree distribution, or more specifically the random variable $K$, which has no counterpart in series system models nor in [15]. Exactly because of this connection through $K$, we are able to show the analytical utility of the present model in the next sections. More specifically, although we are unable to give a closed-form solution to $q$, we manage to conduct analyses in the next three sections to identify topologies (in the sense of degree distributions) and system parameters that lead to more secure systems. These results are valuable because they provide defenders with methods to enhance the security of their systems.

## III. ANALYSIS I: ON THE IMPACT OF VULNERABILITY GRAPH TOPOLOGY

In this section, we analyze the effect of the topology (i.e., degree distribution $D$) of vulnerability graphs topology on $q$. We also discuss its practical significance.

*Theorem 3.1:* If $D$ stochastically increases, then $q$ grows.

Proof of the above theorem is deferred to the Appendix. This theorem is a general result regarding the effect of degree distribution on the security of networked systems. Specifically, it suggests that for any two topologies with $D \preceq_{st} D'$, it holds that $q < q'$, where $q$ (or $q'$) is the probability that a randomly picked vertex is compromised after the system with respect to $D$ (corres. $D'$) enters its steady state. Therefore, it gives an architect an insight for improving the security of an existing system (say, by amending the topology), or a new system by choosing a more appropriate topology.

In order to utilize Theorem 3.1, the system architect or administrator needs to know, for any $D$ and $D'$, whether $D \preceq_{st} D'$ or $D' \preceq_{st} D$. In practice, given any two concrete graphs, statistical degree distributions may be obtained so as to facilitate the comparison. In what follows we consider some general scenarios with respect to three topologies: *regular graphs* with vertices of random degree $D_g$, *random graphs* with vertices of random degree $D_r$, and *power-law graphs* with vertices of random degree $D_{\ell,\nu}$. For a regular graph, denote by $g$ the degree of the vertices. For a random graph, denote by $r$ the edge probability, namely the probability that there is an edge between any pair of vertices. For a power-law graph, the degree follows $\Pr[D_{\ell,\nu} = d] \propto d^{-\nu-1}$, where $\nu \geq 1$ and $\ell \geq 1$ are some positive constants, and $\nu$ is called the power-law exponent.

Equivalently, we let

$$\Pr[D_{\ell,\nu} = d] = \frac{\nu \ell^\nu}{d^{\nu+1}} \quad \text{for } d \geq \ell.$$

First we give a result on the stochastic order of two topologies of the same type. Proof of the following theorem is deferred to the Appendix.

*Theorem 3.2: (stochastic order between two topologies of the same type)* Suppose the parameters are denoted as specified above.

(1) If $g < g'$, then $D_g \preceq_{st} D_{g'}$.

(2) If $r < r'$, then $D_r \preceq_{st} D_{r'}$.

(3) If $\nu < \nu'$, then $D_{\ell,\nu'} \preceq_{st} D_{\ell,\nu}$.

Then we establish the stochastic order of two topologies of different types. Proof of the following theorem is deferred to the Appendix.

*Theorem 3.3: (stochastic order between two topologies of different types)* Suppose the parameters are denoted as specified above.

(1) If $\ell \geq g$, then $D_g \preceq_{st} D_{\ell,\nu}$.

(2) If $\dfrac{\ell^2}{2\pi(|V| - \ell - 1)\nu^2} \leq r \leq \dfrac{\ell}{|V| - 1}$, then $D_r \preceq_{st} D_{\ell,\nu}$.

## A. Practical Significance

Theorems 3.1-3.3 bring the following insights. Given a networked system where the properties of the vertices (via parameters $\alpha$, $\beta$, $\gamma$, and $\eta$) are fixed. Then, the following holds:

(1) When the system's topology is a regular graph, where every node has the same degree, the smaller the vertex degree is, the more secure the system is.

(2) When the system's topology is an Erdos-Renyi random graph, the smaller the edge probability is, the more secure the system is.

(3) When the system's topology is a power-law graph (i.e., the vertex degree follows the power-law distribution), the larger the power-law exponent is (i.e., the lighter the tail of the distribution is), the more secure the system is.

(4) When the system topologies are regular graphs with $\Pr[D_g = g] = 1$ and power-law graphs with $\Pr[D_{\ell,\nu} = d] = \frac{\nu \ell^\nu}{d^{\nu+1}}$ such that $\ell \geq g$, the system with the regular graph topology is more secure.

(5) When the system topologies are random graphs with edge probability $r$ and power-law graphs with $\Pr[D_{\ell,\nu} = d] = \frac{\nu \ell^\nu}{d^{\nu+1}}$ such that $\frac{\ell^2}{2\pi(|V| - \ell - 1)\nu^2} \le r \le \frac{\ell}{|V| - 1}$, the system with the random graph topology is more secure.

*Remark 3.4:* We should mention that Theorem 3.1 does not necessarily imply $C_{t\to\infty}^{(D)} \preceq_{st} C_{t\to\infty}^{(D')}$, where $C_{t\to\infty}^{(D)}$ (or $C_{t\to\infty}^{(D')}$) is the number of compromised vertices in a system with respect to $D$ (corres. $D'$) when the system enters its steady state. This is because $C_{t\to\infty}^{(D)}$ and $C_{t\to\infty}^{(D')}$ depend on both $q$ and $|V|$. However, for two graphs of the same $|V|$, $D \preceq_{st} D'$ does imply $C_{t\to\infty}^{(D)} \preceq_{st} C_{t\to\infty}^{(D')}$, and thus $\mathsf{E}[C_{t\to\infty}^{(D)}] \le \mathsf{E}[C_{t\to\infty}^{(D')}]$, meaning that, on average, there are less compromised vertices in the system corresponding to $D$ than in the system corresponding to $D'$. Finally, it deserves special mention that, since $\mathsf{E}[D] \le \mathsf{E}[D']$ is a necessary, but not sufficient, condition for $D \preceq_{st} D'$, the condition in Theorem 3.1, namely that $D$ stochastically increases, can *not* be substituted by the condition that $\mathsf{E}[D]$ increases.

## IV. Analysis II: On Tuning Parameters

In the previous section we explored the effect of topologies. In this section, we investigate the effect of tuning $\alpha$, $\beta$, $\gamma$, and $\eta$. We also discuss its practical significance.

*Proposition 4.1:* Suppose the parameters are specified as given above. Then, $q$ decreases as $\beta + \eta$ grows, and $q$ increases as $\alpha$ or $\gamma$ grows.

Proof of the above proposition is deferred to the Appendix. The proposition says that security of a system can be improved by either increasing $\beta + \eta$, or decreasing $\alpha$ or $\gamma$. While this can serve as a general guideline for the defenders, an answer to the following question may be of even more value: *Which parameters play more important roles in making the system more secure (i.e., decreasing $q$)?* In order to answer this question, we consider three strategies:

* **Strategy 1**: Increase $\beta$ to $\beta + \omega$, which is equivalent to increase $\beta + \eta$ to $\beta + \eta + \omega$.

* **Strategy 2**: Decrease $\alpha$ to positive $\alpha - \omega$.

* **Strategy 3**: Decrease $\gamma$ to positive $\gamma - \omega$.

We have the following proposition, whose proof is deferred to the Appendix.

*Proposition 4.2:* Suppose we are given $G = (V, E)$, where $E$ is generated according to $D$ that follows a given degree distribution. Suppose the other parameters are specified as given above. Then the following holds:

(1) If $\alpha > \beta + \eta$, then **Strategy 1** is better than **Strategy 2** for $\omega \in (0, \alpha - \beta - \eta]$.

(2) If $\alpha + \gamma\mu < \beta + \eta$, then **Strategy 2** is better than **Strategy 1** for $\omega \in (0, \beta + \eta - \alpha - \gamma\mu]$.

(3) If $\frac{\alpha\mu}{\alpha+\beta+\eta} \geq 1$, then **Strategy 3** is better than **Strategy 2** for $\omega \in (0, \min\{\alpha, \gamma\})$.

### A. Practical significance

The above results are useful. Specifically, Proposition 4.1 serves as a general guideline that shows that security of a system can be improved by either increasing $\beta + \eta$ (e.g., by deploying some reactive security mechanisms such as intrusion detection systems), or decreasing $\alpha$ or $\gamma$ (e.g., by deploying some proactive security mechanisms such as virus filters). Proposition 4.2 gives the conditions under which increasing $\beta$ to $\beta + \omega$, decreasing $\alpha$ to $\alpha - \omega$, or decreasing $\gamma - \omega$ is more effective.

## V. ANALYSIS III: BOUNDING $q$

Since $q$ is determined by Eq. (II.2), which is an implicit equation that we are unable to solve, we instead manage to bound $q$ in this section. As we will see, it is easy to bound $q$ in the case of regular graphs (Proposition 5.1). The case for arbitrary graphs is more involved (Proposition 5.2). Proofs of the propositions are deferred to the Appendix.

*Proposition 5.1: (upper and lower bounding $q$ in regular graphs)* Consider a regular graph $G$ with degree $d$. We have

$$\frac{\alpha}{\alpha + \beta + \eta} \leq q \leq \frac{\alpha + \gamma d}{\alpha + \gamma d + \beta + \eta}. \tag{V.1}$$

Now we upper bound $q$ for arbitrary degree distribution.

*Proposition 5.2: (upper and lower bounding $q$ in arbitrary graphs)* Consider a graph $G$ with an arbitrary degree distribution $D$ with $\mu = \mathsf{E}[D]$. We have

$$\frac{\alpha}{\alpha + \beta + \eta} \leq q \leq \frac{\alpha + \gamma\mu}{\alpha + \beta + \eta + \gamma\mu}.$$

*Corollary 5.3:* Consider a graph $G$ with an arbitrary degree distribution $D$ with $\mu = \mathsf{E}[D]$. Then, in the long run,

$$\frac{|V| \cdot \alpha}{\alpha + \beta + \eta} \leq \mathsf{E}[C_t] \leq \frac{|V| \cdot (\alpha + \gamma\mu)}{\alpha + \beta + \eta + \gamma\mu}. \tag{V.2}$$

### A. Practical Significance

In this subsection we discuss the practical significance of the bounds on $q$ obtained above. Specifically we are interested in deriving some sufficient and necessary conditions under which $C_{t\to\infty} \leq c \cdot |V|$ with a high probability $1 - \varepsilon$, where $0 < c < 1$ may be seen as the threshold of tolerable portion of compromised vertices, and $\varepsilon$ is a given parameter. Since we are unable to derive a closed-form for $q$, the conditions are also based on the bounds of $q$.

Let $\tilde{C}_t$ be the random variable of binomial distribution with parameter $(|V|, \tilde{q})$ and $\hat{C}_t$ be the random variable of binomial distribution with parameter $(|V|, \hat{q})$. Since $\hat{q} \leq q \leq \tilde{q}$, $\hat{C}(t) \preceq_{st} C_t \preceq_{st} \tilde{C}_t$ for large $t$, which implies

$$\Pr[\hat{C}_t \leq x] \geq \Pr[C_t \leq x] \geq \Pr[\tilde{C}_t \leq x], \qquad \text{for all } x. \tag{V.3}$$

First we give a sufficient condition for $\Pr[C_t \leq c\,|V|] \geq 1 - \varepsilon$ in the following proposition, whose proof is deferred to the Appendix.

*Proposition 5.4: (sufficient condition)* If

$$\frac{\alpha + \gamma\mu}{\alpha + \beta + \eta + \gamma\mu} \leq \frac{c^2|V|}{|V| + z^2(\varepsilon)}, \tag{V.4}$$

then $\Pr[C_t \leq c\,|V|] \geq 1 - \varepsilon$.

Now we give a necessary condition for $\Pr[C_t \leq c\,|V|] \geq 1 - \varepsilon$ in the following propostion, whose proof is deferred to the Appendix.

*Proposition 5.5: (necessary condition)* If $\Pr[C_t \leq c\,|V|] \geq 1 - \varepsilon$, then

$$\frac{\alpha}{\alpha + \beta + \eta} \leq \frac{c^2|V|}{|V| + z^2(\varepsilon)}. \tag{V.5}$$

## VI. SIMULATION STUDY

In this section we report the results of our simulation study. The focus is to show the following:

* The stability of $q_i$ and $q$. When the system enters its steady state from the perspective of $q_i$, the probability that node $i \in V$ is compromised, becomes almost stable (i.e., with very small standard deviation). Indeed, once the $q_i$ becomes almost stable, the averaged $q$, the probability that a randomly picked node is compromised, is also stable. This is shown in Section VI-B.

* The impact of graph topology on the underlying system's security. For simplicity, we only considered different topologies of the same type. This is shown in Section VI-C.

* The impact of different parameter tuning methods. For some arbitrarily selected parameters, we compare which methods lead to more secure systems. This is shown in Section VI-D.

* The tightness (or looseness) of the bounds on the number of compromised vertices. This is shown in Section VI-E.

* The validity of the sufficient condition under which the number of compromised nodes is below a threshold with a high probability. This is shown in Section VI-F.

### A. Simulation Setting and Methodology

Recall that our model was built on top of an undirected graph $G = (V, E)$ with no self-loops, where $G$ is certain abstractions of networked systems and every $v \in V$ is initially secure. We considered three different types of topologies: regular graph, random graph, and power-law graph.

In our simulations, we used regular graphs with default node degree value 5, random graphs with default edge probability $p = 0.002$. Regarding power-law graphs, we used two generators: one is the Brite topology generator [18], and the other is the PLOD topology generator [22]. The reason is that while Brite was widely used, it seemingly does not provide us a way to specify the power-law exponents, which is needed for comparing two power-law graphs. This functionality is provided by PLOD. All experiments were performed on graphs with 2000 nodes. Typical parameters were $\alpha = 0.05$, $\beta = 0.2$, and $\gamma = 0.1$, which were used except where otherwise specified. All simulation $C_t$ values were averaged over at least 100 runs, unless otherwise specified. The graphs used in our simulations are summarized in Figure 2.

Recall that our model is a continuous time model. In our simulation, we used an event-driven simulation. An event occurs when a vertex changes state from secure to compromised, or vice-versa. When a vertex's state should change depends on the parameters of $\alpha$, $\beta$, and $\gamma$ as well as the states of its neighbors. For example, if one of a secure vertex's neighbor has just become compromised, a new exponentially-distributed waiting time is calculated by using an updated rate. This can simply replace the existing waiting time due to the *memoryless* property of exponential distributions. In the simulation, time is discretized using a 64-bit floating-point time granularity, and pseudorandom numbers are generated using the Mersenne Twister generator [16].

### B. Stability of the $q_i$'s

In order to show that $q_i$, and thus $q = (\sum_{i \in V} q_i)/|V|$ and $C_t$, will be stable, we measured, for every $i \in V$ in every simulation, $q_i$ from every consecutive interval of length 20, until time step

| Graph type | $|V| = ?$ | Parameter | Average degree $\mu$ |
|---|---|---|---|
| Regular* | 2000 | 5 | 5 |
| Regular | 2000 | (4,3,2) | (4,3,2) |
| Random | 2000 | $p = 0.001$ | 2.235 |
| Random* | 2000 | $p = 0.002$ | 3.957 |
| Random | 2000 | $p = 0.003$ | 6.017 |
| Power-law* (Brite) | 2000 | $m = 2$** | 3.997 |
| Power-law (PLOD) | 2000 | $\nu = 1.421$*** | 3.373 |
| Power-law (PLOD) | 2000 | $\nu = 1.424$*** | 3.356 |
| Power-law (PLOD) | 2000 | $\nu = 1.429$*** | 3.330 |

Fig. 2. Graphs used in our simulations. An asterisk (*) marks graphs that are the default for that type, which are used except where otherwise noted. (**) signifies the following: According to the Brite documentation, this controls the average outdegree. Since we treat the graph as undirected, our average degree is about twice the average outdegree (i.e., average outdegree + average indegree). (***) signifies that this value was computed based on measurement.

330. Then we averaged them over 100 simulation runs. For each vertex, we truncated the first 30 timesteps, then we obtain 15 "samples" of its $q_i$, which allows us to calculate its standard deviation. Based on each vertex's deviation, Figure 3 shows that most vertices have standard deviations less than 3%. Notice that topology has a slight impact on stability. More specifically, for the simulation parameters, regular graphs have a smaller deviation when compared with random graphs, which have a smaller deviation when compared with power-law graphs.

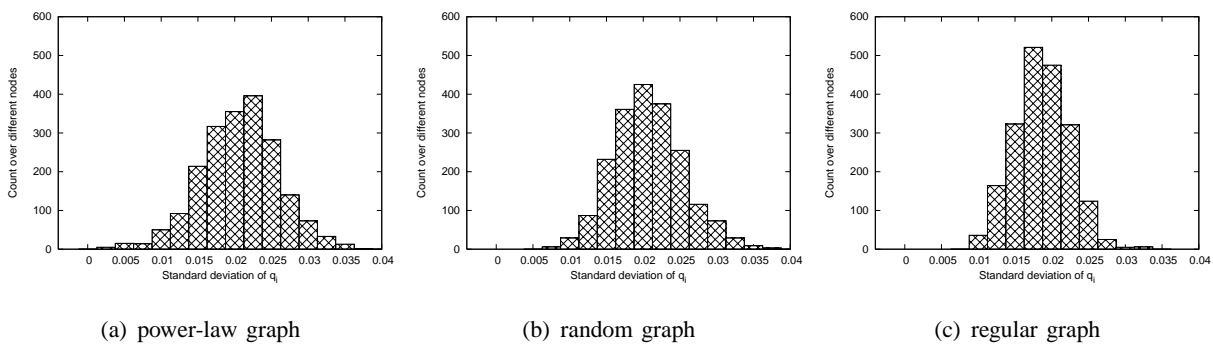

(a) power-law graph                    (b) random graph                    (c) regular graph

Fig. 3. Histogram of standard deviation of $q_i$'s

### C. Impact of Graph Topology on Security

In Section III we showed that the topology of the graph abstracted from a networked system has an impact on security of the system. Figure 4 confirms this. More specifically, Figure 4.(a) shows that regular graphs with lower degree are more secure; Figure 4.(b) shows that random graphs with lower edge probability are more secure; Figure 4.(c) shows that power-law graphs with smaller exponents are more secure.

The most difficult to verify was that power-law graphs with smaller exponents are more secure, as we needed power-law graphs of varying exponent but identical degree. We used the PLOD generator to generate graphs of different exponents. Unfortunately, average degrees of graphs PLOD generated varied considerably, so we generated almost 3000 graphs to try to find three of similar degree and varied exponent. We found three with $\mu \in [3.34, 3.37]$, and $\nu \in [1.421, 1.429]$. These are shown in Figure 4.(c), which is averaged over 2000 runs to show the difference clearly.

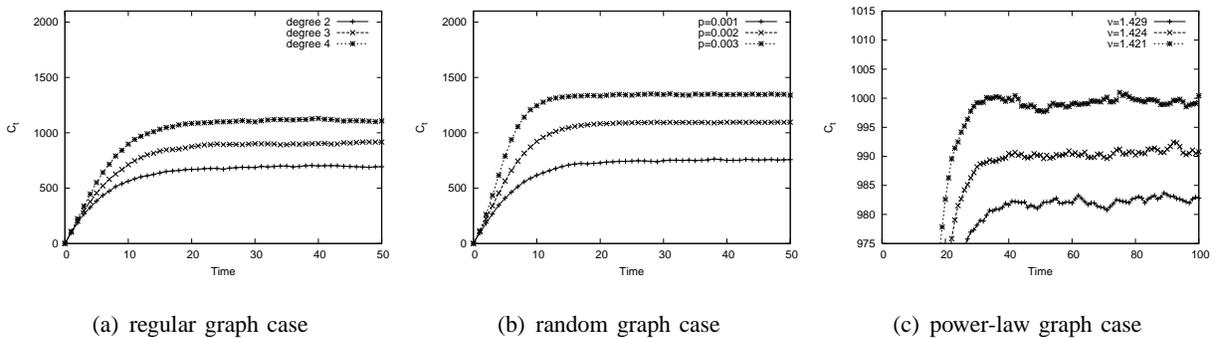

(a) regular graph case            (b) random graph case            (c) power-law graph case

Fig. 4.   Impact of topology on security

### D. Impact of Parameter Tuning

In Section IV we showed tuning parameters in certain way under certain circumstances would lead to more secure systems. Figure 5 confirms this. Specifically, recall that **Strategy 1** is to increase $\beta$ to $\beta + \omega$, **Strategy 2** is to decrease $\alpha$ to $\alpha - \omega$, and **Strategy 3** is to decrease $\gamma$ to $\gamma - \omega$. Figure 5.(a) considers a fixed $G$ as well as parameters $\alpha = 0.1$, $\gamma = 0.1$, $\beta = 0.05$ as the base case, and parameter $\omega = 0.05$. It shows that **Strategy 1** leads to a more secure system than **Strategy 2** does. Figures 5.(b) considers fixed $G$ as well as parameters $\alpha = 0.05$, $\gamma = 0.05$, $\beta = 0.3$ for base case, and parameter $\omega = 0.04$. It shows that **Strategy 2** leads to a more secure system than **Strategy 1** does. Figures 5.(c) considers the base case of fixed $G$ as well as parameters $\alpha = 0.1$, $\gamma = 0.1$, $\beta = 0.2$ for base case, and parameter $\omega = 0.05$. It shows that **Strategy 3** leads to a more secure system than **Strategy 2**,

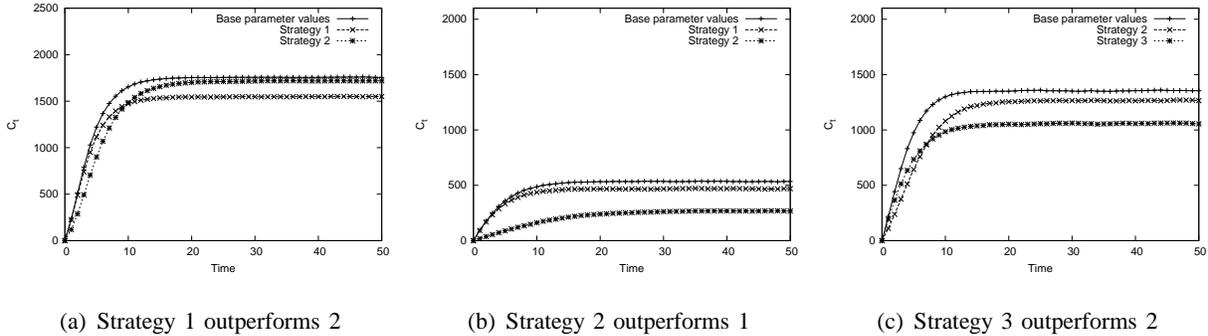

(a) Strategy 1 outperforms 2          (b) Strategy 2 outperforms 1          (c) Strategy 3 outperforms 2

Fig. 5.   On the impact of strategies on security

### E. Accuracy of the Bounds

In Section V we gave upper and lower bounds on $q$, and thus $\mathsf{E}[C_t]$. We would like to confirm that these bounds are valid, and also like to observe how tight (or loose) they are. Since the bounds of $\mathsf{E}[C_t] = q \cdot |V|$, where $q$ depends on parameters, $\alpha$, $\beta$, $\gamma$, $\eta$, and $\nu$, for clarification we set $\eta = 0$ because it has a similar impact as $\beta$ does. For each type of graph $G = (V, E)$, we would like to investigate the impact of varying $\alpha$, $\gamma$, and $\beta$ from 0.1 to 0.5 with step-length 0.1. In order to plot some 3-dimensional surface graphs for $\mathsf{E}[C_t]$, we vary two parameters in each graph with the third one being constant at 0.1. For these graphs, we let gnuplot interpolate the simulation surface from 4 squares by 4 squares to 16 by 16 for viewability. This reduces simulation time from 1600 CPU hours to 100 CPU hours.

**The case of regular graphs**. Figure 6 compares the simulated $C_t$ and the upper and lower bounds of $\mathsf{E}[C_t]$ for a regular graph $G$. Specifically, Figure 6.(a) depicts the effect of $C_t$ while varying $\alpha$ and $\beta$, Figure 6.(b) depicts the effect of $C_t$ while varying $\alpha$ and $\gamma$, Figure 6.(c) depicts the effect of $C_t$ while varying $\gamma$ and $\beta$.

From them we draw the following observations:

1) For regular graphs with fixed $\gamma = 0.1$, when $\alpha$ is significantly greater than 0.1 and $\beta$ is significantly less than 0.5, the upper bound is indeed very tight. It seems that the smaller

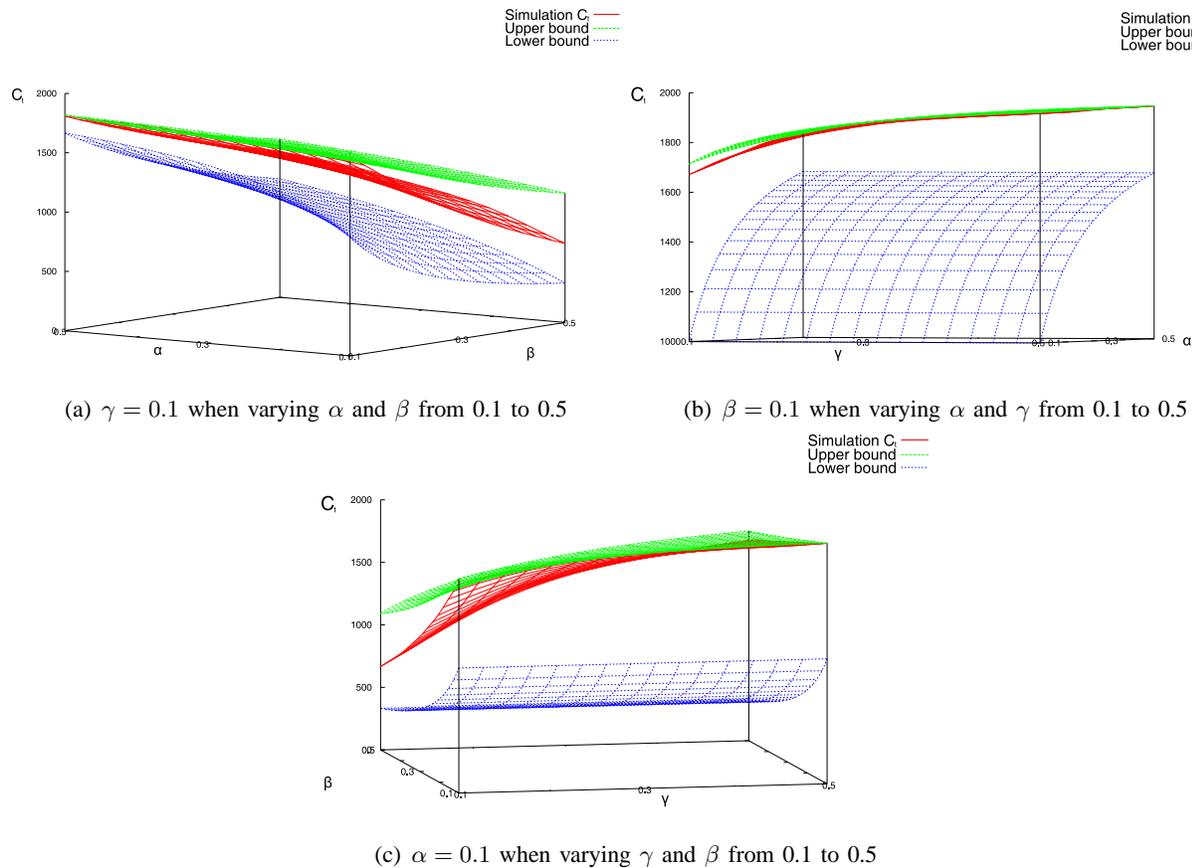

(a) $\gamma = 0.1$ when varying $\alpha$ and $\beta$ from 0.1 to 0.5

(b) $\beta = 0.1$ when varying $\alpha$ and $\gamma$ from 0.1 to 0.5

(c) $\alpha = 0.1$ when varying $\gamma$ and $\beta$ from 0.1 to 0.5

Fig. 6.   Accuracy of the upper and lower bounds in the case of regular graphs ($|V| = 2,000$): simulation results vs. bounds

$\beta - \alpha$ the tighter the upper bound (including the case $\alpha > \beta$). The cause is perhaps that before the compromised vertices successfully launch attacks against other vertices, they have been detected and appropriately dealt with. On the other hand, the lower bound is constantly loose. The reason is perhaps that it does not capture $\gamma$, which reflects the interactions between the vertices.

2) For regular graphs with fixed $\beta = 0.1$, the upper bound is always tight except when both $\alpha$ and $\gamma$ approach 0. On the other hand, the lower bound is always loose, and becomes meaningless when $\alpha$ approaches 0.1. The reason is perhaps that it does not capture $\gamma$, which which reflects the interactions between the vertices.

3) For regular graphs with fixed $\alpha$, the upper bound is always tight except when $\gamma$ approaches 0.1 and $\beta$ approaches 0.5. The cause is unclear. On the other hand, the lower bound is always loose, but does not oscillate much. The reason is perhaps that it does not capture $\gamma$, which reflects the interactions between the vertices.

In summary, the upper bound is quite tight, especially in the case of fixed $\beta$, whereas the lower bound is quite loose in most cases. Moreover, it seems the tightness of the upper bounds is sensitive to $\beta$. This means we can use the upper bounds as a good approximation in further reasoning (e.g., the sufficient condition under which the compromised number of nodes is below a threshold with a high probability; see below).

**The case of random graphs**. Figure 7 compares the simulated $C_t$ and the upper and lower bounds of $\mathsf{E}[C_t]$ for a random graph $G$. Specifically, Figure 7.(a) depicts the effect of $C_t$ while varying $\alpha$ and $\beta$, Figure 7.(b) depicts the effect of $C_t$ while varying $\alpha$ and $\gamma$, Figure 7.(c) depicts the effect of $C_t$ while varying $\gamma$ and $\beta$.

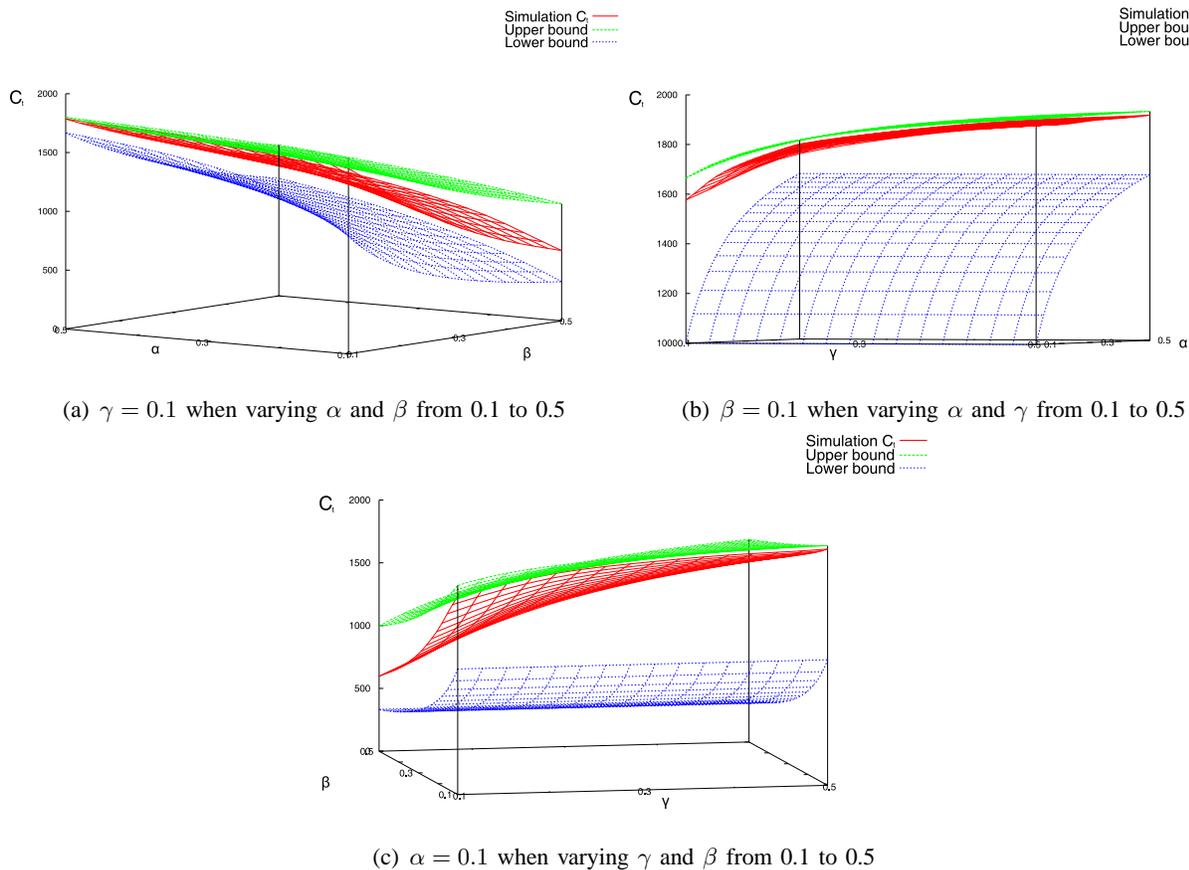

(a) $\gamma = 0.1$ when varying $\alpha$ and $\beta$ from 0.1 to 0.5

(b) $\beta = 0.1$ when varying $\alpha$ and $\gamma$ from 0.1 to 0.5

(c) $\alpha = 0.1$ when varying $\gamma$ and $\beta$ from 0.1 to 0.5

Fig. 7.   Accuracy of the upper and lower bounds in the case of random graphs ($|V| = 2,000$): simulation results vs. bounds

From them we draw the following observations:

1) For random graphs with fixed $\gamma = 0.1$, when $\alpha$ is significantly greater than 0.1 and $\beta$ is significantly less than 0.5, the upper bound is indeed very tight. It seems that the smaller $\beta - \alpha$ the tighter the upper bound (including the case $\alpha > \beta$). The cause is perhaps that before the compromised vertices successfully launch attacks against other vertices, they have been detected and appropriately dealt with. On the other hand, the lower bound is constantly loose. The reason is perhaps that it does not capture $\gamma$, which reflects the interactions between the vertices.

2) For random graphs with fixed $\beta = 0.1$, the upper bound is always tight except when both $\alpha$ and $\gamma$ approach 0. On the other hand, the lower bound is always loose, and becomes meaningless when $\alpha$ approaches 0.1. The reason is perhaps that it does not capture $\gamma$, which reflects the interactions between the vertices.

3) For random graphs with fixed $\alpha$, the upper bound is always tight except when $\gamma$ approaches 0.1 and $\beta$ approaches 0.5. The cause is unclear. On the other hand, the lower bound is always loose, but does not oscillate much. The reason is perhaps that it does not capture $\gamma$, which reflects the interactions between the vertices.

In summary, the upper bound is quite tight, especially in the case of fixed $\beta$, whereas the lower bound is quite loose in most cases. Moreover, it seems the tightness of the upper bounds is sensitive to $\beta$. This means we can use the upper bounds as a good approximation in further reasoning (e.g., the sufficient condition under which the compromised number of nodes is below a threshold with a high probability; see below). Notice that there is no significant difference between the above regular graph case and the above random graph one.

**The power-law graph case**. Figure 8 compares the simulated $C_t$ and the upper and lower bounds of $\mathsf{E}[C_t]$ for a power-law graph $G$. Specifically, Figure 8.(a) depicts the effect of $C_t$ while varying $\alpha$ and $\beta$, Figure 8.(b) depicts the effect of $C_t$ while varying $\alpha$ and $\gamma$, Figure 8.(c) depicts the effect of $C_t$ while varying $\gamma$ and $\beta$.

From them we draw the following observations:

1) For power-law graphs with fixed $\gamma = 0.1$, when $\alpha$ is significantly greater than 0.1 and $\beta$ is significantly less than 0.5, the upper bound is indeed very tight. It seems that the smaller $\beta - \alpha$ the tighter the upper bound (including the case $\alpha > \beta$). The cause is perhaps that before the compromised vertices successfully launch attacks against other vertices,

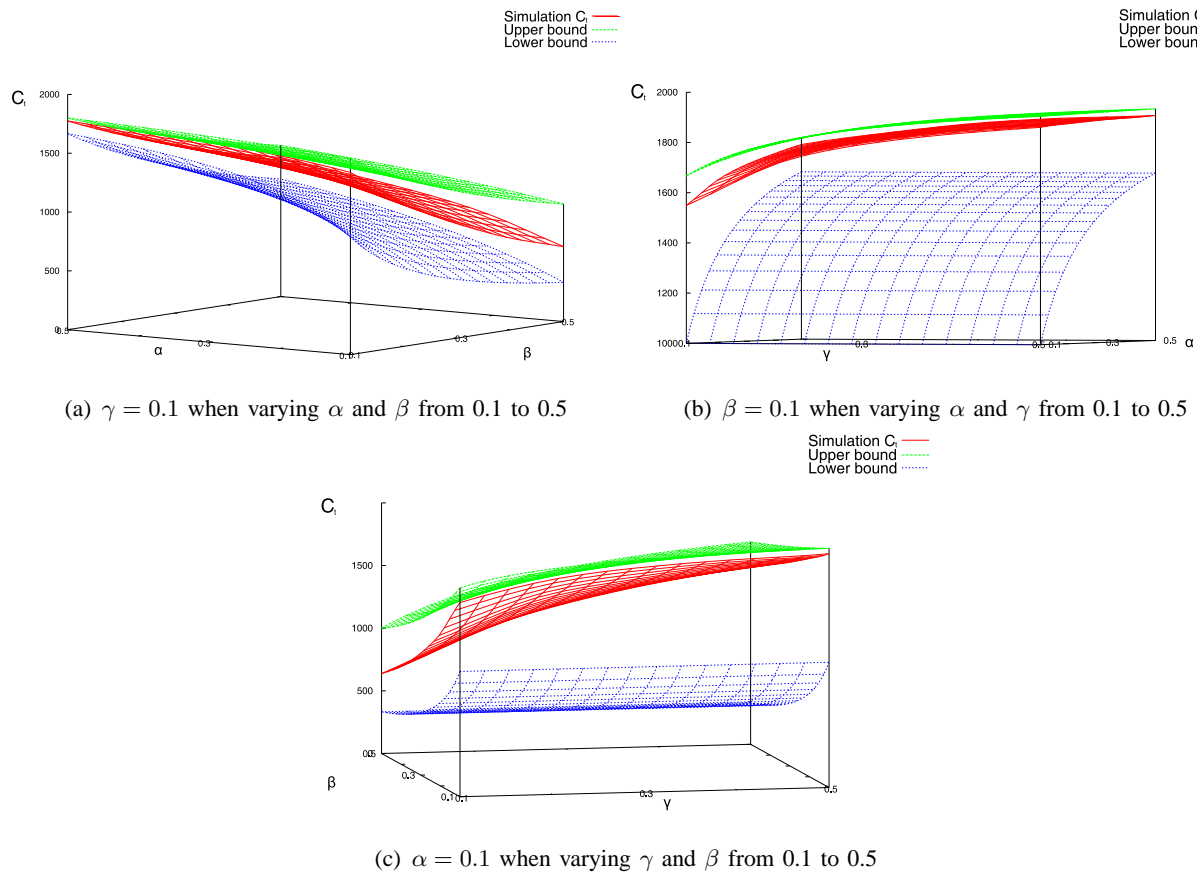

(a) $\gamma = 0.1$ when varying $\alpha$ and $\beta$ from 0.1 to 0.5

(b) $\beta = 0.1$ when varying $\alpha$ and $\gamma$ from 0.1 to 0.5

(c) $\alpha = 0.1$ when varying $\gamma$ and $\beta$ from 0.1 to 0.5

Fig. 8.   Accuracy of the upper and lower bounds in the case of power-law graphs ($|V| = 2,000$): simulation results vs. bounds

they have been detected and appropriately dealt with. On the other hand, the lower bound is constantly loose. The reason is perhaps that it does not capture $\gamma$, which reflects the interactions between the vertices.

2) For power-law graphs with fixed $\beta = 0.1$, the upper bound is somewhat tight except when both $\alpha$ and $\gamma$ approach 0. On the other hand, the lower bound is always loose, and becomes meaningless when $\alpha$ approaches 0.1. The reason is perhaps that it does not capture $\gamma$, which reflects the interactions between the vertices.

3) For power-law graphs with fixed $\alpha$, the upper bound is always tight except when $\gamma$ approaches 0.1 and $\beta$ approaches 0.5. The cause is unclear. On the other hand, the lower bound is always loose, but does not oscillate much. The reason is perhaps that it does not capture $\gamma$, which reflects the interactions between the vertices.

In summary, the upper bound is somewhat tight, especially in the case of fixed $\beta$, whereas the lower bound is quite loose in most cases. Moreover, it seems the tightness of the upper bound is sensitive to $\beta$. This means we can still use the upper bounds as a good approximation in further reasoning. Notice that there is a significant difference between the above regular / random graph cases and the power-law graph one.

**Summary**. The lower bound given be our model is always loose. The upper bound given by our model is often quite tight, and the tightness varies insignificantly from regular graphs (with degree 5), random graphs (with average degree $0.002 \times 2000 = 4$), to power-law graphs (with average degree 3.997). This similarity may have been caused by the fact that the average degrees of the graphs are about the same. In each type of graph, it seems that the tightness of the upper bounds is sensitive to $\beta$.

### F. On the Validity of the Sufficient Condition

The above simulation results indicated that the upper bound given by our model is quite tight. Now we further validate the utility of the upper bounds for the purpose of specifying a sufficient condition under which the number of compromised vertices is below a threshold (e.g., one third) with a high probability, namely $\Pr[C_t \leq c \, |V|] \geq 1 - \varepsilon$. Such a sufficient condition is useful because it further simplifies the task of system administrators in tuning parameters towards a given goal.

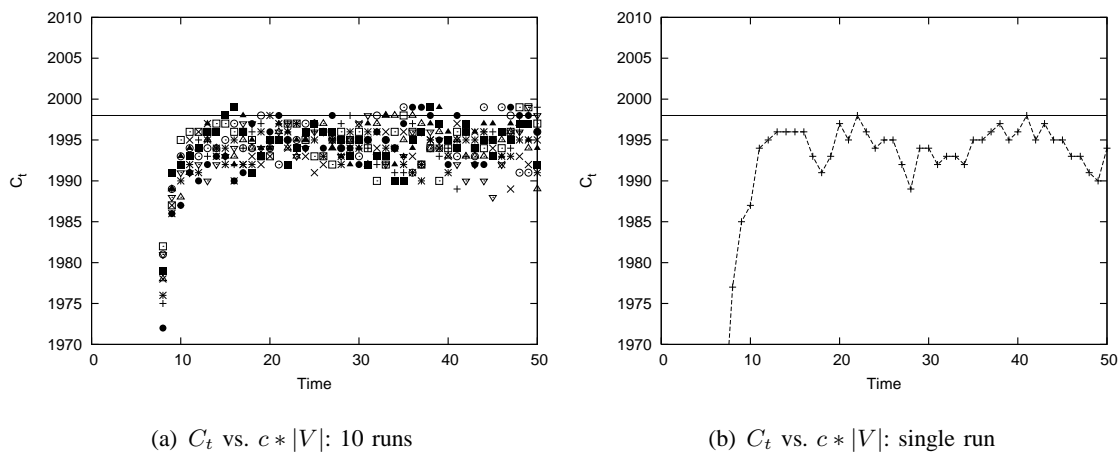

(a) $C_t$ vs. $c * |V|$: 10 runs                               (b) $C_t$ vs. $c * |V|$: single run

Fig. 9.   On the validity of the sufficient condition

For this purpose, we tested a number of cases including many where the premise "marginally" held; in no case were we able to find more than $\varepsilon$ instances where $C_t > c * |V|$ – usually there were no instances at all. Figure 9.(a) uses a scatter graph of $C_t$ from 10 simulation runs to depict the most excess simulation values of $C_t$ we found, using parameters $\alpha = 0.25$, $\gamma = 0.1$, and $\beta = 0.002$. Note that instances of $C_t$ above the line $c \cdot |V|$ are much fewer than $\varepsilon = 0.159 = 15.9\%$, where $c = 0.5$, $\varepsilon = 0.159$, and $z = 2$. For a even better visual effect, Figure 9.(b) plots $C_t$ vs. $c \cdot |V|$ in a single simulation run (randomly selected).

## VII. RELATED WORK

There have been some attempts to understand security of networked systems by considering a system as a whole. Nevertheless, our approach can address some unique and important questions.

**Existing epidemic-like models vs. our model**. Epidemic models [17], [14], [6], [4] have been adopted to investigate the spreading of computer viruses or worms pioneered by [13]. As mentioned before, our model can be seen as a generalization of traditional epidemic models because traditional models often (the exceptions are discussed below) assume that every node or vertex has equal contact with everyone else (i.e., effectively regular graphs), and that the rate of infection is largely determined by the density of the infected individuals. Such homogeneous models cannot capture the heterogeneity of the vertex degrees in vulnerability graphs. Whereas, our model can.

The aforementioned exceptions that considered non-homogeneous graphs are due to the following. Epidemic spreading in the Barabasi-Albert power-law networks [7] is investigated in [19], [24], [25], [23]. More general non-homogeneous graphs are investigated in [34], [33], [11]. The differences between these models and ours are the following.

- All of the models in [34], [33], [11] can *only* capture the spreading behavior of attacks, and cannot capture the fact that computers can get compromised because of their own reasons (e.g., a user downloads and executes a malicious code). In contrast, our model captures this (through the parameter $\alpha > 0$).

- None of the models in [34], [33], [11] can answer questions our model deals with, namely "How a defender should "tune" system configurations or parameters so as to improve security." This is true even though [33] indicated that indicates that CAN [28] might be more secure than Chord [31] against attacks, because the former is a lower degree regular graph.

**Privilege graph and attack graph based approaches vs. our approach**. In a *privilege graph* [9], [21], a vertex represents a set of privileges on some objects and an arc represents a vulnerability. An arc exists from one vertex to another if there is a method allowing a user owning the former vertex's privileges to obtain those of the latter. In an *attack graph*, a vertex represents the state of a network (i.e., the values assigned to relevant system attributes such as specific vulnerabilities on various hosts and connectivity between hosts), and an edge represents a step in an attack (cf. [27], [12], [3] and the references therein). A designated vertex (or set of vertices) represents the initial state(s), and each transition represents a specific exploit that an attack can carry out.

The main difference between privilege and attack graphs and ours lies in the model *purpose* and *scalability*. From a purpose perspective, privilege graphs can be used to estimate the effort an attacker, in order to defeat the system security objectives, might expend to exploit the vulnerabilities. Attack graphs can be used to identify end-to-end attack paths by chaining together the vulnerabilities uncovered by the vulnerability scanners. In contrast, our model focuses on investigating the impact of system attributes, such as topologies and vertex properties, on the security of the system. Moreover, our approach suggests methods to "tune" the parameters to lead to more secure systems. From a scalability perspective, privilege and attack graph based approaches suffer from limited *scalability*, because of their inherent exponential state explosion [3], [8]. In contrast, our approach is scalable because there is no issue of state explosion.

**Key challenge graph based approach vs. our approach**. In a *key challenge graph* [8], a vertex represents a host, and an arc represents a *key challenge* — an abstraction to capture access control. A key challenge is, for instance, a password authentication prior to accessing to a resource. The starting point of an attack could be one or more vertices, which are assumed to be initially in the control of the attacker. The target of an attack could be one or more vertices, for which the attacker knows the location and the paths to reach them. A successful attack is a sequence of zero or more vertices not in the initial set but eventually containing all the target vertices. The cost of an attack is measured as the sum of the effort required to compromise individual vertices by attempting to counter the key challenges on the edges. Since there are multiple paths from the starting point to the target, a problem of particular interest is to find an attack path of minimum cost.

The main difference between the key challenge graph based approach and ours lies in the model *purpose* and *capability*. From a purpose perspective, the key challenge graph based approach emphasizes an algorithmic aspect of finding an optimal attack path, namely that the adversary can achieve its goal with minimal effort or cost. As noted before, our model focuses on the impact of system attributes on the security of the system, and suggests methods to "tune" the attributes to lead to more secure systems. From a capability perspective, the key challenge graph based approach can only capture the attack behaviors with respect to some specific starting points (i.e., the vertices initially compromised); otherwise, it will encounter the exponential explosion problem. In contrast, our modeling approach does not need to know the initially compromised vertices. Instead, it can accommodate the scenario of no initially compromised vertices.

**Connectivity-oriented analysis vs. our security-oriented analysis**. Connectivity-oriented analysis of networked systems can be traced back to the early days of random graph theory [10]. A central problem in this context is to investigate the network reliability subject to edge removals, namely the probability that graph remains connected after removing some edges. Recently, such analysis has been extended to explore the impact of topologies of complex communication networks, including the impact of removing vertices, which is more damaging because removal of a vertex implies the removal of all its edges as well. It turns out that there is a strong correlation between connectivity and network topology [1]. For example, consider networks that have the same number of vertices and edges, and differ only in their degree distributions. Then, power-law networks are more robust than random networks against random vertex failures, but are more vulnerable when the most connected vertices are targeted [2].

Connectivity-oriented analysis does not necessarily bring much insight for security analysis. This is because a malicious attacker would be more likely to use the compromised vertices as "stepping stones" to attack some more vertices, rather than simply undermining the network connectivity. Therefore, the connectivity of a system may never be jeopardized, but the security has been undermined. Our model focuses on the security aspect by taking into consideration the perspective of topologies as well as vertex properties. Moreover, our model suggests methods to "tune" the parameters to lead to more secure systems.

## VIII. Conclusion

We presented a novel modeling approach to investigating security of networked systems based on their vulnerability graph abstractions. Our model carries insights into designing more secure new systems or enhancing security of existing systems. In particular, it offers methods for tuning system configurations and parameters so as to make systems more secure.

While we believe that this paper moves a significant step towards full-fledged quantitative security analyses, we hope that it will inspire more investigations in tackling this challenging problem. For example, our trick of accommodating network topology through its degree distribution possibly represents an approach to fulfilling the program "from dependability to security" envisioned in [20].

**Acknowledgement**. We thank Raj Boppana for his assistance in conducting our simulation study. We thank the anonymous reviewers for their comments that helped improve the paper.

This work was supported in part by ARO, NSF, AFOSR, UTSA CIAS, and UTSA TRAC Award. The views and conclusions contained in the article are those of the authors and should not be interpreted as, in any sense, the official policies or endorsements of the government or the agencies.

## References

[1] R. Albert and A. Barabasi. Statistical mechanics of complex networks. *Reviews of Modern Physics*, 74:47–97, 2002.

[2] R. Albert, H. Jeong, and A. Barabasi. Error and attack tolerance of complex networks. *Nature*, 406:378482, 2000.

[3] P. Ammann, D. Wijesekera, and S. Kaushik. Scalable, graph-based network vulnerability analysis. In *Proceedings of the 9th ACM conference on Computer and communications security (CCS'02)*, pages 217–224, 2002.

[4] R. Anderson and R. May. *Infectious Diseases of Humans*. Oxford University Press, 1991.

[5] The Computing Research Association. Four grand challenges in trustworthy computing. `http://www.cra.org/Activities/grand.challenges/`, 2003.

[6] N. Bailey. *The Mathematical Theory of Infectious Diseases and Its Applications*. 2nd Edition. Griffin, London, 1975.

[7] A. Barabasi and R. Albert. Emergence of scaling in random networks. *Science*, 286:509–512, 1999.

[8] R. Chinchani, A. Iyer, H. Ngo, and S. Upadhyaya. Towards a theory of insider threat assessment. In *2005 International Conference on Dependable Systems and Networks (DSN'05)*, pages 108–117, 2005.

[9] M. Dacier and Y. Deswarte. Privilege graph: an extension to the typed access matrix model. In *Proc. of ESORICS'94*, volume 875 of *Lecture Notes in Computer Science*, pages 319–334. Springer, 1994.

[10] P. Erdos and A. Renyi. On the evolution of random graphs. *Publications of the Mathematical Institute of the Hungarian Academy of Sciences*, 5:17–61, 1960.

[11] A. Ganesh, L. Massoulie, and D. Towsley. The effect of network topology on the spread of epidemics. In *Proceedings of IEEE Infocom 2005*, 2005.

[12] S. Jha and J. Wing. Survivability analysis of networked systems. In *Proceedings of the 23rd International Conference on Software Engineering (ICSE'01)*, pages 307–317, Washington, DC, USA, 2001. IEEE Computer Society.

[13] J. Kephart and S. White. Directed-graph epidemiological models of computer viruses. In *IEEE Symposium on Security and Privacy*, pages 343–361, 1991.

[14] W. Kermack and A. McKendrick. A contribution to the mathematical theory of epidemics. *Proc. of Roy. Soc. Lond. A*, 115:700–721, 1927.

[15] B. Madan, K. Goševa-Popstojanova, K. Vaidyanathan, and K. Trivedi. A method for modeling and quantifying the security attributes of intrusion tolerant systems. *Perform. Eval.*, 56(1-4):167–186, 2004.

[16] M. Matsumoto and T. Nishimura. Mersenne twister: a 623-dimensionally equidistributed uniform pseudo-random number generator. *ACM Trans. Model. Comput. Simul.*, 8(1):3–30, 1998.

[17] A. McKendrick. Applications of mathematics to medical problems. *Proc. of Edin. Math. Soceity*, 14:98–130, 1926.

[18] A. Medina, A. Lakhina, I. Matta, and J. Byers. Brite: An approach to universal topology generation. In *Proc. International Symposium in Modeling, Analysis and Simulation of Computer and Telecommunication Systems (MASCOTS'01)*, pages 346–356, 2001.

[19] Y. Moreno, R. Pastor-Satorras, and A. Vespignani. Epidemic outbreaks in complex heterogeneous networks. *European Physical Journal B*, 26:521–529, 2002.

[20] D. Nicol, W. Sanders, and K. Trivedi. Model-based evaluation: From dependability to security. *IEEE Trans. Dependable Secur. Comput.*, 1(1):48–65, 2004.

[21] R. Ortalo, Y. Deswarte, and M. Kaaniche. Experimenting with quantitative evaluation tools for monitoring operational security. *IEEE Trans. Softw. Eng.*, 25(5):633–650, 1999.

[22] C. Palmer and J. Steffan. Generating network topologies that obey power laws. In *GLOBECOM*, 2000.

[23] R. Pastor-Satorras and A. Vespignani. Epidemics and immunization in scale-free networks. *Handbook of Graphs and Networks: From the Genome to the Internet*.

[24] R. Pastor-Satorras and A. Vespignani. Epidemic dynamics and endemic states in complex networks. *Physical Review E*, 63:066117, 2001.

[25] R. Pastor-Satorras and A. Vespignani. Epidemic dynamics in finite size scale-free networks. *Physical Review E*, 65:035108, 2002.

[26] M. Pease, R. Shostak, and L. Lamport. Reaching agreement in the presence of faults. *J. ACM*, 27(2):228–234, 1980.

[27] C. Phillips and L. Swiler. A graph-based system for network-vulnerability analysis. In *Proceedings of the 1998 workshop on New security paradigms (NSPW'98)*, pages 71–79, New York, NY, USA, 1998. ACM Press.

[28] S. Ratnasamy, P. Francis, M. Handley, R. Karp, and S. Schenker. A scalable content-addressable network. In *Proceedings of ACM SIGCOMM'01*, pages 161–172, 2001.

[29] S. Ross. *Stochastic Processes*. Wiley Series in Probability and Mathematical Statistics. John Wiley & Sons, Inc, 1996.

[30] M. Shaked and J. Shanthikumar. *Stochastic Orders and Their Applications*. Academic Press, San Diego (CA), 1994.

[31] I. Stoica, R. Morris, D. Karger, M. Kaashoek, and H. Balakrishnan. Chord: A scalable peer-to-peer lookup service for internet applications. In *Proceedings of ACM SIGCOMM'01*, pages 149–160, 2001.

[32] K. Trivedi. *Probability and Statistics with Reliability, Queuing and Computer Science Applications*. John Wiley and Sons, 2001.

[33] J. Wang, L. Lu, and A. Chien. Tolerating denial-of-service attacks using overlay networks – impact of topology. In *Proceedings of ACM workshop on survivable and self-regnerative systems*, 2003.

[34] Y. Wang, D. Chakrabarti, C. Wang, and C. Faloutsos. Epidemic spreading in real networks: An eigenvalue viewpoint. In *Proc. of the 22nd IEEE Symposium on Reliable Distributed Systems (SRDS'03)*, pages 25–34, 2003.

## APPENDIX

The following is the proof of Theorem 3.1.

*Proof:* Suppose $D'$ is another random degree such that $D \preceq_{st} D'$. In parallel to $K$ being the distribution of the number of a vertex's compromised neighbors with respect to $D$, we denote by $K'$ the number of a randomly picked vertex's compromised neighbors with respect to $D'$. Recall that $q$ is the probability that a randomly picked vertex is compromised with respect to $D$. If we treat $K'$ as the binomial distribution with parameters $(d', q)$ given $D' = d'$. Since

$$K = \sum_{i=1}^{D} A_i \text{ and } K' = \sum_{i=1}^{D'} A_i,$$

where the $A_i$'s are independently and identically distributed Bernoulli random variable with parameter $q$, it is straightforward to see that $K \preceq_{st} K'$. Since $\frac{\beta + \eta}{\alpha + \gamma x}$ decreases as $x$ grows, we have

$$\mathsf{E}\left[\frac{\beta + \eta}{\alpha + \gamma K}\right] \geq \mathsf{E}\left[\frac{\beta + \eta}{\alpha + \gamma K'}\right],$$

that is

$$h(\alpha, \beta, \gamma, \eta, D; x) \geq h(\alpha, \beta, \gamma, \eta, D'; x).$$

This means that, for any $x \in [0, 1]$, $h(\alpha, \beta, \gamma, \eta, D; x)$ decreases as $D$ stochastically increases.

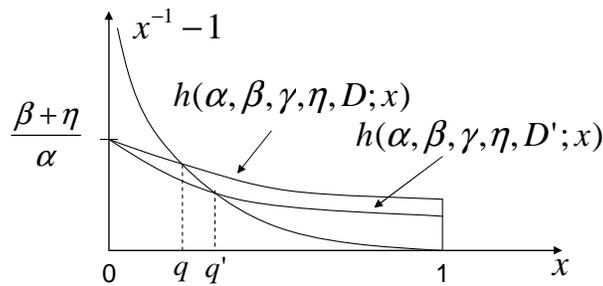

Fig. 10.   Illustration of $h(\alpha, \beta, \gamma, \eta, D; x)$

Note that both $\frac{1}{x} - 1$ and $h(\alpha, \beta, \gamma, \eta, D; x)$ strictly decrease in interval $(0, 1)$ as $x$ grows. Moreover, $\frac{1}{x} - 1 - h(\alpha, \beta, \gamma, \eta, D; x) > 0$ as $x \to 0$ and $\frac{1}{x} - 1 - h(\alpha, \beta, \gamma, \eta, D; x) < 0$ as $x \to 1$. Therefore, Eq. (II.1) must have exactly one real root in $[0, 1]$. Since $q$ can be seen as the point of intersection of the two curves $\frac{1}{x} - 1$ and $h(\alpha, \beta, \gamma, \eta, D; x)$, as illustrated in Figure 10, $D \preceq_{st} D'$ implies $q < q'$. Therefore, $q$ grows as $D$ stochastically increases. ∎

The following is the proof of Theorem 3.2.

*Proof:* (1) We notice that

$$\Pr[D_g > d] = \begin{cases} 1, & d < g, \\ 0, & d \geq g, \end{cases} \quad \text{and } \Pr[D_{g'} > d] = \begin{cases} 1, & d < g', \\ 0, & d \geq g'. \end{cases}$$

Since $g < g'$, it is clear that $\Pr[D_g > d] \leq \Pr[D_{g'} > d]$ for $d = 0, 1, 2, \ldots, |V| - 1$. That is, $D_g \preceq_{st} D_{g'}$.

(2) For the random graph with respect to degree distribution $D_r$, let $B_1, B_2, \ldots, B_{|V|-1}$ be the respective indicators whether there is an edge between a randomly picked vertex and any other vertex. Definition of random graphs implies

$$\Pr[B_i = 1] = r = 1 - \Pr[B_i = 0] \quad \text{for } i = 1, \ldots, |V| - 1.$$

In parallel, for the random graph with respect to degree distribution $D_{r'}$ we have

$$\Pr[B_i' = 1] = r' = 1 - \Pr[B_i' = 0] \text{ for } i = 1, \ldots, |V| - 1.$$

Then

$$D_r = \sum_{i=1}^{|V|-1} B_i \quad \text{and} \quad D_{r'} = \sum_{i=1}^{|V|-1} B_i'.$$

Since $r \leq r'$, $B_i \preceq_{st} B_i'$ for $i = 1, 2, \ldots, |V| - 1$, it holds that

$$\sum_{i=1}^{|V|-1} B_i \preceq_{st} \sum_{i=1}^{|V|-1} B_i'.$$

That is, $D_r \preceq_{st} D_{r'}$.

(3) Since

$$\Pr[D_{\ell,\nu} > d] = \begin{cases} 1, & d < \ell, \\ \frac{\ell^\nu}{d^\nu}, & d \geq \ell, \end{cases} \quad \text{and } \Pr[D_{\ell,\nu'} > d] = \begin{cases} 1, & d < \ell, \\ \frac{\ell^{\nu'}}{d^{\nu'}}, & d \geq \ell, \end{cases}$$

$\nu' > \nu$ implies

$$\frac{\ell^\nu}{d^{\nu'}} \leq \frac{\ell^\nu}{d^\nu},$$

and hence $\Pr[D_{\ell,\nu} > d] \geq \Pr[D_{\ell,\nu'} > d]$. That is, $D_{\ell,\nu'} \preceq_{st} D_{\ell,\nu}$.                                ■

The following is the proof of Theorem 3.3.

*Proof:* For (1), we notice that

$$\Pr[D_g > d] = \begin{cases} 1, & d < g, \\ 0, & d \geq g, \end{cases} \quad \text{and} \quad \Pr[D_{\ell,\nu} > d] = \begin{cases} 1, & d < \ell, \\ \dfrac{\ell^\nu}{d^\nu}, & d \geq \ell. \end{cases}$$

Hence $\ell \geq g$ implies

$$\Pr[D_{\ell,\nu} > d] \geq \Pr[D_g > d], \qquad \text{for all } d \geq 0,$$

and thus $D_g \preceq_{st} D_{\ell,\nu}$.

For (2), we notice that

$$\Pr[D_{\ell,\nu} = d] = \frac{\nu \ell^\nu}{d^{\nu+1}}, \quad \text{for } d = \ell, \cdots, |V| - 1.$$

By Laplace's theorem, for any $d = 0, 1, \cdots, |V| - 1$,

$$\begin{aligned} \Pr[D_r = d] &= \binom{|V|-1}{d} r^d (1-r)^{|V|-d-1} \\ &\approx \frac{1}{\sqrt{2\pi \cdot (|V|-1) \cdot r(1-r)}} \exp\left\{ -\frac{(d - r \cdot (|V|-1))^2}{2 \cdot (|V|-1) \cdot r(1-r)} \right\}. \end{aligned}$$

If $r \leq \dfrac{\ell}{|V|-1}$ then

$$\mathsf{E}[D_r] = (|V|-1)r \leq \ell, \tag{.1}$$

and $1 - r \geq \dfrac{|V| - \ell - 1}{|V| - 1}$. If $\dfrac{\ell^2}{2\pi(|V|-\ell-1)\nu^2} \leq r$, then,

$$r(1-r) \geq (1-r)\frac{\ell^2}{2\pi(|V|-\ell-1)\nu^2} \geq \frac{\ell^2}{2\pi \cdot (|V|-1) \cdot \nu^2},$$

which is equivalent to

$$\frac{1}{\sqrt{2\pi \cdot (|V|-1) \cdot r(1-r)}} \leq \frac{\nu}{\ell}.$$

Thus, we have, for all $d = 0, 1, \cdots, |V| - 1$,

$$\frac{1}{\sqrt{2\pi \cdot (|V|-1) \cdot r(1-r)}} \exp\left\{ -\frac{(d - (|V|-1) \cdot r)^2}{2 \cdot (|V|-1) \cdot r(1-r)} \right\} \leq \frac{\nu}{\ell}. \tag{.2}$$

In combination with (.1) and (.2), it holds that, for any $d = \ell, \ell+1, \cdots, |V| - 1$,

$$\Pr[D_r = d] \leq \Pr[D_{\ell,\nu} = \ell].$$

Therefore, it holds that for all $d = 0, 1, 2, \cdots, |V| - 1$,

$$\Pr[D_r > d] \leq \Pr[D_{\ell,\nu} > d],$$

and thus $D_r \preceq_{st} D_{\ell,\nu}$.                                                                                 ∎

The following is the proof of Proposition 4.1.

*Proof:* On one hand, Eq. (II.1) shows that for given $D$ and $x$, $h(\alpha, \beta, \gamma, \eta, D; x)$ increases as $\beta + \eta$ grows, and decreases as $\alpha$ or $\gamma$ grows. On the other hand, as shown in Figure 10, we observe (1) both $\frac{1}{x} - 1$ and $h(\alpha, \beta, \gamma, \eta, D; x)$ decrease with respect to $x \in (0, 1)$, and (2) $q$ is the crossing point of the two curves $\frac{1}{x} - 1$ and $h(\alpha, \beta, \gamma, \eta, D; x)$. Therefore, it can be concluded that $q$ decreases as $\beta + \eta$ grows, and $q$ increases as $\alpha$ or $\gamma$ grows.                                                                                 ∎

The following is the proof of Proposition 4.2.

*Proof:* For given $D$, denote by $q_1$ the point of intersection of the curves $x^{-1} - 1$ and $h(\alpha, \beta + \omega, \gamma, \eta, D; x)$, by $q_2$ the point of intersection of the curves $x^{-1} - 1$ and $h(\alpha - \omega, \beta, \gamma, \eta, D; x)$, and by $q_3$ the point of intersection of the curves $x^{-1} - 1$ and $h(\alpha - \omega, \beta, \gamma - \omega, \eta, D; x)$.

(1) For our purpose, it suffices to verify whether $q - q_1 \geq q - q_2$ or $q - q_2 > q - q_1$. Since

$$\min\{h(\alpha, \beta + \omega, \gamma, \eta, D; x), h(\alpha - \omega, \beta, \gamma, \eta, D; x)\} > h(\alpha, \beta, \gamma, \eta, D; x),$$

it is equivalent to check which strategy leads to a larger cross point with curve $x^{-1} - 1$. Let

$$
\begin{aligned}
f(D, q) &= h(\alpha, \beta + \omega, \gamma, \eta, D; x) - h(\alpha - \omega, \beta, \gamma, \eta, D; x) \\
&= \mathsf{E}\left[\frac{\beta + \eta + \omega}{\alpha + \gamma K}\right] - \mathsf{E}\left[\frac{\beta + \eta}{\alpha - \omega + \gamma K}\right] \\
&= \mathsf{E}\left[\frac{\omega(\alpha + \gamma K) - \omega(\beta + \eta) - \omega^2}{(\alpha + \gamma K)(\alpha - \omega + \gamma K)}\right] \\
&= \omega \cdot \mathsf{E}\left[\frac{\alpha + \gamma K - \beta - \eta - \omega}{(\alpha + \gamma K)(\alpha - \omega + \gamma K)}\right] \\
&\geq \omega \cdot \mathsf{E}\left[\frac{\alpha + \gamma K - \beta - \eta - \omega}{(\alpha + \gamma \cdot |V|)(\alpha - \omega + \gamma \cdot |V|)}\right].
\end{aligned}
$$

If $\alpha \geq \beta + \eta$ and $\omega \in (0, \alpha - \beta - \eta]$, then $f(D, q) > 0$, or $q_1 < q_2$. This means that increasing $\beta$ leads to a more secure system than decreasing $\alpha$ for $\omega \in (0, \alpha - \beta - \eta]$. That is, **Strategy 1** is better than **Strategy 2**.

(2) Suppose $\alpha + \gamma\mu < \beta + \eta$ and $\omega \in (0, \beta + \eta - \alpha - \gamma\mu]$. Then,

$$
\begin{aligned}
f(D, q) &= \omega \cdot \mathsf{E}\left[\frac{\alpha + \gamma K - \beta - \eta - \omega}{(\alpha + \gamma K)(\alpha - \omega + \gamma K)}\right] \\
&\leq \frac{\omega}{\alpha(\alpha - \omega)}\mathsf{E}\left[\alpha + \gamma K - \beta - \eta - \omega\right] \\
&\leq \frac{\omega(\alpha + \gamma\mu - \beta - \eta - \omega)}{\alpha(\alpha - \omega)} \\
&\leq 0.
\end{aligned}
$$

This means that $q_1 \geq q_2$, and thus **Strategy 2** is better than **Strategy 1**.

(3) Notice that

$$
\mathsf{E}[K] = \sum_{d=1}^{|V|-1} \mathsf{E}[K|D = d] \cdot \Pr[D = d] = \sum_{d=1}^{|V|-1} qd \cdot \Pr[D = d] = q\mathsf{E}[D] = q\mu.
$$

Since

$$
h(\alpha, \beta, \gamma, \eta, D; x) = \mathsf{E}\left[\frac{\beta + \eta}{\alpha + \gamma K}\right] \leq \frac{\beta + \eta}{\alpha},
$$

the root of Eq. (II.1) is always greater than or equal to the root of $\frac{1}{x} - 1 = \frac{\beta + \eta}{\alpha}$. That is, $q \leq \frac{\alpha}{\alpha + \beta + \eta}$.

For $0 < \omega < \min\{\alpha, \gamma\}$, if $\frac{\alpha\mu}{\alpha + \beta + \eta} \geq 1$, then $\mathsf{E}[K] \geq 1$. As a result,

$$
\begin{aligned}
&h(\alpha, \beta, \gamma - \omega, \eta, D; x) - h(\alpha - \omega, \beta, \gamma, \eta, D; x) \\
&= \mathsf{E}\left[\frac{\beta + \eta}{\alpha + (\gamma - \omega)K}\right] - \mathsf{E}\left[\frac{\beta + \eta}{\alpha + \gamma K - \omega}\right] \\
&= (\beta + \eta) \cdot \mathsf{E}\left[\frac{\omega(K - 1)}{[\alpha + (\gamma - \omega)K][\alpha + \gamma K - \omega]}\right] \\
&\geq \frac{(\beta + \eta)\omega(\mathsf{E}[K] - 1)}{[\alpha + (\gamma - \omega)(|V| - 1)][\alpha + \gamma(|V| - 1) - \omega]} \\
&\geq 0.
\end{aligned}
$$

Hence, $q_3 < q_2$, and **Strategy 3** is better than **Strategy 2**.                    ∎

The following is the proof of Proposition 5.1.

*Proof:* For a regular graph $G$ with degree $D \equiv d$,

$$
\frac{\beta + \eta}{\alpha + \gamma d} \leq \mathsf{E}\left[\frac{\beta + \eta}{\alpha + \gamma K}\right] \leq \frac{\beta + \eta}{\alpha}.
$$

By Eq. (II.2), it holds that

$$
\frac{1}{1 + \frac{\beta + \eta}{\alpha}} \leq q = \frac{1}{1 + \mathsf{E}\left[\dfrac{\beta + \eta}{\alpha + \gamma K}\right]} \leq \frac{1}{1 + \frac{\beta + \eta}{\alpha + \gamma d}},
$$

and hence the proposition holds.                                                          ∎

The following is the proof of Proposition 5.2.

*Proof:*  (Upper bound) By Jessen's inequality,

$$\mathsf{E}\left[\frac{1}{\alpha + \gamma K}\right] \geq \frac{1}{\alpha + \gamma q \mu} \geq \frac{1}{\alpha + \gamma \mu}.$$

Then,

$$q = \frac{1}{1 + \mathsf{E}\left[\dfrac{\beta + \eta}{\alpha + \gamma K}\right]} \leq \tilde{q} \overset{def}{=} \frac{1}{1 + \dfrac{\beta + \eta}{\alpha + \gamma \mu}} = \frac{\alpha + \gamma \mu}{\alpha + \beta + \eta + \gamma \mu}.$$

(Lower bound) Consider a random variable $K_0$ such that

$$\Pr[K_0 = 0] = \Pr[K = 0] = \sum_{d=0}^{N-1} (1 - q)^d \Pr[D = d] = \mathsf{E}[(1-q)^D],$$

and

$$\Pr[K_0 = 1] = 1 - \mathsf{E}[(1-q)^D].$$

It is obvious that $K_0 \preceq_{st} K$. Note that

$$\frac{1}{\alpha + \gamma K} \preceq_{st} \frac{1}{\alpha + \gamma K_0},$$

we have

$$\mathsf{E}\left[\frac{1}{\alpha + \gamma K}\right] \leq \mathsf{E}\left[\frac{1}{\alpha + \gamma K_0}\right].$$

Since

$$\begin{aligned}
\mathsf{E}\left[\frac{1}{\alpha + \gamma K_0}\right] &= \frac{1}{\alpha}\mathsf{E}[(1-q)^D] + \frac{1}{\alpha + \gamma}\left(1 - \mathsf{E}[(1-q)^D]\right) \\
&= \frac{1}{\alpha + \gamma} + \left(\frac{1}{\alpha} - \frac{1}{\alpha + \gamma}\right)\mathsf{E}[(1-q)^D] \\
&\leq \frac{1}{\alpha + \gamma} + \frac{\gamma}{\alpha(\alpha + \gamma)} \\
&= \frac{1}{\alpha},
\end{aligned}$$

it follows that

$$q = \frac{1}{1 + \mathsf{E}\left[\dfrac{\beta + \eta}{\alpha + \gamma K}\right]} \geq \frac{1}{1 + \mathsf{E}\left[\dfrac{\beta + \eta}{\alpha + \gamma K_0}\right]} \geq \hat{q} \overset{def}{=} \frac{1}{1 + \dfrac{\beta + \eta}{\alpha}} = \frac{\alpha}{\alpha + \beta + \gamma}.$$

The following is the proof of Proposition 5.4.                                          ∎

*Proof:* Since the size of the graph $(V, E)$ is usually large, by using *Laplace*'s theorem, we have

$$\Pr[\tilde{C}_t \leq c\,|V|] \approx \Phi\left(\frac{c\,|V| - \tilde{q}\,|V|}{\sqrt{\tilde{q}(1-\tilde{q})\,|V|}}\right),$$

where $\Phi$ is the distribution function of the standard normal random variable.

For any small $\varepsilon \in (0,1)$, let $z(\varepsilon)$ be the $1 - \varepsilon$ quantile of the standard normal distribution. Set $\Phi\left(\frac{c\,|V| - \tilde{q}\,|V|}{\sqrt{\tilde{q}(1-\tilde{q})\,|V|}}\right) \geq 1 - \varepsilon$, we have $\frac{c\,|V| - \tilde{q}\,|V|}{\sqrt{\tilde{q}(1-\tilde{q})\,|V|}} \geq z(\varepsilon)$. Denote by

$$f_1(\tilde{q}) = \tilde{q}^2\left(|V| + z^2(\varepsilon)\right) - \tilde{q}\left(2c|V| + z^2(\varepsilon)\right) + c^2|V| \geq 0. \tag{.3}$$

Since $f_1(c) < 0$, $f_1(0) > 0$, and $f_1(1) > 0$, $f_1(\tilde{q}) = 0$ has two real solutions in the unit interval $(0,1)$ and $c$ lies between these two solutions. Notice that $\tilde{q} = 1$ must be excluded (otherwise, Eq. (.3) does not hold). This implies that we only need to consider the smaller root of $f_1(\tilde{q}) = 0$, which is

$$\begin{aligned}
\lambda(\varepsilon, c, V) &= \frac{2c|V| + z^2(\varepsilon) - \sqrt{z^4(\varepsilon) + 4c|V|(1-c)z^2(\varepsilon)}}{2\left[|V| + z^2(\varepsilon)\right]} \\
&\geq \frac{2c|V| + z^2(\varepsilon) - \left[z^2(\varepsilon) + 2c|V|(1-c)\right]}{2\left[|V| + z^2(\varepsilon)\right]} \\
&= \frac{c^2|V|}{|V| + z^2(\varepsilon)},
\end{aligned}$$

by taking Eq. (V.3) into consideration, for any $\varepsilon \in (0,1)$, if

$$\frac{\alpha + \gamma\mu}{\alpha + \beta + \eta + \gamma\mu} \leq \frac{c^2|V|}{|V| + z^2(\varepsilon)},$$

then $\Pr[C_t \leq c\,|V|] \geq \Pr[\tilde{C}_t \leq c\,|V|] \geq 1 - \varepsilon$.                     ■

The following is the proof of Proposition 5.5.

*Proof:* By using *Laplace*'s theorem,

$$\Pr[\hat{C}_t \leq c\,|V|] \approx \Phi\left(\frac{c\,|V| - \hat{q}\,|V|}{\sqrt{\hat{q}(1-\hat{q})\,|V|}}\right).$$

For any small $\varepsilon \in (0,1)$, $\Phi\left(\frac{c\,|V| - \hat{q}\,|V|}{\sqrt{\hat{q}(1-\hat{q})\,|V|}}\right) \geq 1 - \varepsilon$ is equivalent to

$$f_2(\hat{q}) = \hat{q}^2\left(|V| + z^2(\varepsilon)\right) - \hat{q}\left(2c|V| + z^2(\varepsilon)\right) + c^2|V| \geq 0.$$

Then, for large $t$, $\Pr[\hat{C}_t \leq c\,|V|] \geq 1 - \varepsilon$ implies $\hat{q} \leq \frac{c^2|V|}{|V| + z^2(\varepsilon)}$. In combination with Eq. (V.3), we have, if $\Pr[C_t \leq c\,|V|] \geq 1 - \varepsilon$ then $\Pr[\hat{C}_t \leq c\,|V|] \geq 1 - \varepsilon$ and hence Eq. (V.5).     ■

Xiaohu Li is a Research Associate in the Department of Computer Science, University of Texas at San Antonio. He is also a full professor in the School of Mathematics and Statistics, Lanzhou University, China. His research interests include aging property, stochastic order, sample spacing, $k$-out-of-$n$ structure, and non-parametric life-testing. He obtained PhD in 2000 from Lanzhou University, China.

Timothy Paul Parker received a B.S. in computer science from Baylor University and an M.S. from Rice University. Subsequently he worked in industry for five years, first as a kernel developer at Hewlett-Packard and then doing natural language processing R&D at a government contractor. He is now a Ph.D. student in computer security at the University of Texas at San Antonio. His interests include the intersections of cryptography and systems.

Shouhuai Xu is an Assistant Professor in the Department of Computer Science, University of Texas at San Antonio. His expertise and research interests include cryptography and quantitative security modeling and analysis of networked systems. He earned his PhD from Fudan University, China. More information about his research can be found at `www.cs.utsa.edu/~shxu`.